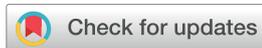

Check for updates

Cite this: DOI: 10.1039/d3cp03088a

# Theoretical insights into the structural, electronic and thermoelectric properties of the inorganic biphenylene monolayer†

Ajay Kumar,  Parbati Senapati  and Prakash Parida  *

Being motivated by a recently synthesized biphenylene carbon monolayer (BPN), using first principles methods, we have studied its inorganic analogue (B–N analogue) named I-BPN. A comparative study of structural, electronic and mechanical properties between BPN and I-BPN was carried out. Like BPN, the stability of I-BPN was verified in terms of formation energy, phonon dispersion calculations, and mechanical parameters (Young's modulus and Poisson's ratio). The chemical inertness of I-BPN was also investigated by adsorbing an oxygen molecule in an oxygen-rich environment. It has been found that the B–B bond favours the oxygen molecule to be adsorbed through chemisorption. The lattice transport properties reveal that the phonon thermal conductivity of I-BPN is ten times lower than that of BPN. The electronic band structure reveals that I-BPN is a semiconductor with an indirect bandgap of 1.88 eV, while BPN shows metallic behaviour. In addition, we investigated various thermoelectric properties of I-BPN for possible thermoelectric applications. The thermoelectric parameters, such as the Seebeck coefficient, show the highest peak value of 0.00289 V K$^{-1}$ at 300 K. Electronic transport properties reveal that I-BPN is highly anisotropic along the $x$ and $y$-axes. Furthermore, the thermoelectric power factor as a function of chemical potential shows a peak value of 0.057 W m$^{-1}$ K$^{-2}$ along the $x$-axis in the p-type doping region. The electronic figure of merit shows a peak value of approximately unity. However, considering lattice thermal conductivity, the peak value of the total figure of merit ($ZT$) reduces to 0.68(0.46) for p-type and 0.56(0.48) for n-type doping regions along the $x(y)$ direction at 900 K. It is worth noting that our calculated $ZT$ value of I-BPN is higher than that of many other reported B–N composite materials.

## 1. Introduction

Graphene is a monolayer of sp$^2$-hybridized carbon atoms arranged in a honeycomb structure. It is one of the earliest fascinating two-dimensional materials, famous for its linear dispersion relation. It extended a new era in the study of atomically thin-layered materials.[1–4] After the experimental confirmation of graphene by the exfoliation technique in 2005, additional two-dimensional single entity layers, such as silicene,[5] germanene,[6] borophene,[7,8] phosphorene,[9] and arsenene[10] and hetero-structures, such as GaAs,[11] hexagonal boron-nitride[12] (h-BN), and transition metal dichalcogenides,[13] have been studied. In the meantime, h-BN draws attention to theoretical and experimental studies.[12,14,15] h-BN is a graphene-like honeycomb structure. In graphene, the rhombus unit cell

Department of Physics, Indian Institute of Technology Patna, Bihta, Bihar, India.
E-mail: pparida@iitp.ac.in

† Electronic supplementary information (ESI) available. See DOI: https://doi.org/10.1039/d3cp03088a

has identical carbon atoms at the 1/3 and 2/3 positions of its longer diagonal, whereas in h-BN, the same locations are occupied by a boron (B) atom and a nitrogen (N) atom. Moreover, h-BN is iso-electronic with graphene because both accommodate 12 electrons per unit cell. Despite the structural similarities, the electronic band structures of both systems are very different. Graphene is a semi-metal Dirac material, whereas h-BN is an insulator with a large bandgap of 5.6 eV.[14] The binary composition of h-BN distinguishes it from carbon allotropes, and the bond between the boron and nitrogen atoms is partially ionic due to electronegativity differences. This partial charge transfer from B to N could disclose a significant variation in the band structure.[16] Furthermore, h-BN is an intriguing material for two-dimensional systems because of its chemical inertness, good insulating properties for use as a substrate to frame thin films, and thermal stability mixed with mechanical robustness.[17–19] Also, because of its broad bandgap range and atomically ordered thickness, h-BN has recently been described as a complementary metal-oxide semiconductor (CMOS), the most reliable gate insulator in low-dimensional material-based transistors. In addition, combining stacked graphene and h-BN heterostructure in energy storage devices has been reported to improve the accessible surface area and ion storage capacity.[20,21] Moreover, graphene, h-BN, and their heterogeneous structures have been reported from magnetic application and spintronics points of view.[22] Further, the piezoelectric voltage coefficient of a single BN nanoflake (NF) is being studied, and the energy harvesting capabilities of 2D h-BN NF-based flexible piezoelectric energy devices are reported.[23] There are huge numbers of works that try to explore various BN-based structures for different application visions.

In this work, being inspired by a recently synthesized ultra-flat biphenylene carbon monolayer (BPN) by Fan *et al.* *via* a bottom-up approach, we have proposed an inorganic 2D biphenylene network of B–N, named I-BPN (shown in Fig. 1).[24] BPN is composed of regularly spaced four, six, and eight-membered rings of sp[2] carbon atoms. Yunhao *et al.* and Seunghan *et al.* individually reported the modulation of the electronic properties of BPN by varying the concentration of hydrogen and halogens at different sites.[25,26] Bafekry *et al.* reported the electronic and dielectric properties of BPN using the first principles calculations.[27] The inorganic counterpart of graphene is the hexagonal boron nitride sheet, often called h-BN, which shares an identical honeycomb lattice arrangement found in graphene. As a result, h-BN can produce carbon-like nanostructures such as porous structures,[28,29] closed cage 22 structures,[30] and nanotubes.[31] Although the B–N composition networks of carbon allotropes show different electronic properties, these are stuck in their geometrical and mechanical stability with carbon-based structures. The stable carbon allotrope crystals and their analogous networks for boron–nitrogen composition have been studied extensively. For example, graphenylene, a unique porous network of non-delocalized carbon atoms, was reported,[32] and its B–N composed network called inorganic-graphenylene was studied.[33] The B–N composition of other 2D carbon allotrope networks like T-carbon and graphyne was also studied.[34,35]

In light of these factors, a comparative study between BPN and I-BPN has been done regarding their structural stability, mechanical strength and electronic properties. The electronic band structure of BPN shows metallic behaviour with tilted Dirac cones near the Fermi energy.[27] Such types of tilted Dirac cones are generally found in Weyl materials.[36] Its metallic nature is due to delocalized $p_z$ electrons over the 2D biphenylene network. Therefore, BPN may not be a good thermoelectric candidate because a finite narrow gap is required for better thermoelectric features. On the other hand, the B–N bond is partially ionic due to the localization of charges on the atomic sites. For illustration, graphene has covalently bonded carbon atoms that lead to its semi-metallic nature, which in turn makes it a poor thermoelectric material. The h-BN sheet also faces a similar fate of poor thermoelectric behaviour because of its wide gap insulating nature due to the ionic character of B–N bonds. The proposed I-BPN structure is expected to have a narrow band gap due to the presence of partially ionic bonds in the structure. Hence, compared to BPN, exploring the electronic properties of I-BPN will be interesting, which may show better thermoelectric features due to its narrow band gap. The electronic thermoelectric parameters such as the Seebeck coefficient, electrical conductivity, thermopower and figure of merit have been explored using semi-classical Boltzmann transport theory.

Further, thermal conductivity plays an important role in exploring the thermoelectric material. Generally, low (especially phonon) thermal conductivity is highly appreciable for thermoelectric characteristics. In addition, the binary composition structure shows low phonon thermal conductivity compared to the single entity structure.[37,38] BPN consists of carbon atoms only, and its phonon thermal conductivity at room temperature is reported to be of the order of $\sim 10^2$ W m$^{-1}$ K$^{-1}$.[39] It is expected that the I-BPN should have lower phonon thermal conductivity than BPN because of its binary composition of two different entities, boron and nitrogen atoms. Therefore, exploring the phonon thermal conductivity of I-BPN would also be interesting for thermoelectric features.

## 2. Computational details

BPN and I-BPN structures have been studied in the Density Functional Theory (DFT) framework using the Vienna Ab initio Simulation Package (VASP). It used the Projector Augmented Wave (PAW) method, which plays an important role in interactions between the valence electrons and the core electrons with periodic boundary conditions. The pseudopotentials of B, C and N are estimated for electronic configurations having valence electrons $2s^2 2p^1$, $2s^2 2p^2$ and $2s^2 2p^3$, respectively. We have considered the generalized gradient approximation (GGA) for exchange–correlation potential formulated by Perdew–Burke–Ernzerhof (PBE). GGA-PBE stands out as a commonly used functional among various exchange–correlation functionals for different 2D materials (mainly composed of atoms of low atomic numbers) due to its ability to balance computational accuracy and cost-keeping without much bargaining the qualitative conclusions. However, it is worth noting that GGA-PBE can sometimes underestimate the electronic bandgap. In contrast, alternative functionals like hybrid HSE and the GW approximation can provide more precise bandgap predictions at the expense of higher computational resources. GGA-PBE has been safely used by many groups and proved not to be a bad choice for exploring physical and chemical properties of similar 2D materials, like graphene,[40] graphyne,[41] BN monolayer,[42] borophene[43] and many other systems.

For choosing a reasonable energy cutoff and *k*-point grid, a few convergence tests have been carried out and the corresponding results are plotted in Fig. S1 in the ESI.† A convergence test for plane wave energy cutoff over an energy range of 100 to 1000 eV has been done. As shown in Fig. S1 (ESI†), the energy cutoff value beyond 450 eV converges the total energy to a reasonable accuracy. Thus, we choose mesh cutoff = 600 eV to maintain a balance between the accuracy in calculation and low computational time. Similarly, the $k$-point grid is used to divide the first Brillouin zone for calculating the energy value of the system. A convergence test for $k$-mesh has been done by considering a $k$-point grid ranging from $1 \times 1 \times 1$ to $35 \times 32 \times 1$. Although the total energy of our system gets converged for a $k$-point grid of $12 \times 9 \times 1$ (see Fig. S1, ESI†), we still choose the grid size of $25 \times 21 \times 1$ to be on the safe side. The proposed crystal structures are fully relaxed with the force tolerance value of $10^{-3}$ eV Å$^{-1}$ per atom by a conjugate-gradient algorithm. Further, the energy tolerance throughout the calculation is $10^{-8}$ eV. Both structures have rectangular unit cells, which follow the periodic condition in the $x$–$y$ plane. A vacuum of 20 Å is given to avoid interaction along the $z$-axis. The electronic band calculation has been performed for both monolayers along the high symmetry points $\Gamma$–$X$–$S$–$Y$–$\Gamma$.

We have performed the phonon dispersion calculation to show the dynamic stability of the BPN and I-BPN monolayers. For dynamic studies, a supercell is generally required to capture the long-wavelength fluctuations in atomic movements in the periodic structure. The lattice constant of the supercell greater than 10 Å is enough to capture most of the longer wavelengths associated with bonds between two atoms, as C–C/B–N bond lengths are less than 2 Å in our systems. In our calculations, we have utilized a supercell of dimension $4 \times 3 \times 1$, where the approximate lattice constants are $a = 15(15)$ Å and $b = 14(13)$ Å for the BPN (I-BPN) structure. This supercell size is chosen to effectively accommodate all long-wavelength vibrations of both structures during *ab initio* molecular dynamics (AIMD) and phonon dispersion calculations. We have used the Phonopy package and VASP to calculate the force sets and force constants for phonon calculation. For AIMD simulation and phonon calculations, a supercell size of $4 \times 3 \times 1$ and a $k$-mesh of $9 \times 11 \times 1$ have been used in I-BPN monolayers. As a reciprocal cell is squeezed due to a supercell in the direct lattice, a less fine $k$-mesh is sufficient for $k$-point sampling. A Nose thermostat *NVT* canonical ensemble has been employed with a time step of 1 fs over 5000 fs at two different temperatures, 300 K and 600 K.

In order to study the mechanical properties, the generalized Hooke's law is given as

$$\sigma_{ij} = C_{ijkl}\epsilon_{kl}; \ C_{ijkl} = \frac{1}{2}\frac{\partial^2 U}{\partial\epsilon_{ij}\partial\epsilon_{kl}} \ (\text{where } i, j, k, l = 1, 2, 3) \quad (1)$$

where $\sigma$ and $\epsilon$ are the second-rank stress and strain tensors, and $C$ is the fourth-rank stiffness constant. These fourth-order stiffness constants are further represented as $C_{ijkl} \rightarrow C_{pq}$ ($p,q = 1,2,3,4,5,6$) in Voigt notation.[44] For illustration $ij \rightarrow p$ is given as $11 \rightarrow 1$, $22 \rightarrow 2$, $33 \rightarrow 3$, $21 = 12 \rightarrow 4$, $13 = 31 \rightarrow 5$, $23 = 32 \rightarrow 6$.

The elastic strain energy per unit area can be expressed as

$$U(\epsilon_{11}, \epsilon_{22}) = \frac{1}{2}C_{11}\epsilon_{11}{}^2 + C_{22}\epsilon_{11}{}^2 + C_{12}\epsilon_1\epsilon_2 + \ldots \quad (2)$$

Here, $\epsilon_{11}$ and $\epsilon_{22}$ are the strains along the $x$- and $y$-axes, and $C_{11}$, $C_{22}$, $C_{44}$ and $C_{12}$ are stiffness constants in Voigt notation.

Lattice transport properties such as phonon thermal conductivity ($\kappa_{ph}$) have been calculated using the ShengBTE package.[45] A supercell of $4 \times 3 \times 1$ and a $k$-mesh of $9 \times 11 \times 1$ have been used in the calculations for both BPN and I-BPN monolayers to find the lattice thermal conductivity. The Phonopy package has been used to obtain second-order force constants, and the symmetric displacements are used to calculate the forces required for dynamical matrices. The same pseudopotentials and plane-wave basis cutoff energy have been used along with a $9 \times 11 \times 1$ $k$-point grid. The third-order anharmonic interatomic force constants (IFCs) are calculated by creating a $4 \times 3 \times 1$ supercell of the BPN and I-BPN. The third-order IFC considers interactions up to four nearest neighbours. The $4 \times 3 \times 1$ supercell has generated 1618 displacement datasets for both BPN and I-BPN with an atomic displacement of 0.01 Å. Further, the second and third-order IFCs have been used as input to the ShengBTE package for solving the linearised phonon Boltzmann transport equation. A dense $120 \times 120 \times 1$ $k$-mesh is used for the calculation of $\kappa_{ph}$. $\kappa_{ph}$ has been obtained by single-mode relaxation time approximation (RTA) within ShengBTE[45] using the following equation:

$$\kappa_{ph}^{\alpha\beta} = \frac{1}{K_B T^2 \Omega N}\sum_\lambda f_0(f_0+1)(\hbar\omega_\lambda)^2 \eta_\lambda^\alpha F_\lambda^\beta \quad (3)$$

where $\Omega$ is the volume of the unit cell, $N$ is the number of $q$ points uniformly distributed over the Brillouin zone, $\omega_\lambda$ is the angular frequency of phonon modes, $\eta_\lambda^\alpha$ is the phonon group velocity along the $\alpha$ direction, $f_0$ is the Bose–Einstein distribution function, and $F_\lambda^\beta$ is the projection of the mean free displacement along the $\beta$ direction. Furthermore, non-analytical corrections have been applied to the force constants for phonon dispersion and related calculations, such as Born effective charge and dielectric constant.

The electronic transport properties have been investigated using semi-classical Boltzmann transport equations with energy-independent relaxation time and rigid band approximations, as implemented in the BoltzTraP program. The following equations can be used to express the thermoelectric-related variables (in terms of the tensor), such as electrical conductivity ($\sigma_{\alpha\beta}$), conductivity related to the thermal gradient, and electronic thermal conductivity ($k_{\alpha\beta}^0$) along the $\alpha$ and $\beta$ directions:

$$\sigma_{\alpha\beta}(T;\mu) = \frac{1}{\Omega}\int \sigma_{\alpha\beta}(\varepsilon)\left[-\frac{\partial f_\mu(T;\varepsilon)}{\partial\varepsilon}\right]d\varepsilon \quad (4)$$

$$v_{\alpha\beta}(T;\mu) = \frac{1}{eT\Omega}\int \sigma_{\alpha\beta}(\varepsilon)(\varepsilon-\mu)\left[-\frac{\partial f_\mu(T;\varepsilon)}{\partial\varepsilon}\right]d\varepsilon \quad (5)$$

$$k_{\alpha\beta}^0(T;\mu) = \frac{1}{e^2\Omega T}\int \sigma_{\alpha\beta}(\varepsilon)(\varepsilon-\mu)^2\left[-\frac{\partial f_\mu(T;\varepsilon)}{\partial\varepsilon}\right]d\varepsilon \quad (6)$$

The Seebeck coefficient ($S_{\alpha\beta}$) can be easily calculated by using these tensors quantities,

$$S_{\alpha\beta} = \sum_{\gamma} (\sigma^{-1})_{\alpha\gamma} v_{\beta\gamma} \qquad (7)$$

where $T$, $\Omega$, $\mu$ and $f$ are the absolute temperature, cell volume, chemical potential, and Fermi–Dirac distribution, respectively. $\sigma_{\alpha\beta}(\varepsilon)$ represents the density of state energy projected conductivity tensor, which is expressed by

$$\sigma_{\alpha\beta}(\varepsilon) = \frac{1}{N} \sum_{i,k} \sigma_{\alpha\beta}(i,k) \delta(\varepsilon - \varepsilon_{i,k}) \qquad (8)$$

where $N$ represents the number of $k$-points, $\varepsilon_{i,k}$ represents electron-band energies (band index $i$) and $\sigma_{\alpha\beta}(i,k)$ denotes the conductivity tensor, which is as follows,

$$\sigma_{\alpha\beta}(i,k) = e^2 \tau_{i,k} \vartheta_\alpha(i,k) \vartheta_\beta(i,k) \qquad (9)$$

where $e$ is the charge of the electron, $\tau_{i,k}$ is the relaxation time, $\vartheta_\alpha(i,k)$ and $\vartheta_\beta(i,k)$ are group velocities expressed as $\vartheta_\alpha(i,k) = \frac{1}{\hbar} \frac{\partial \varepsilon_{i,k}}{\partial k_\alpha}$, $\vartheta_\beta(i,k) = \frac{1}{\hbar} \frac{\partial \varepsilon_{i,k}}{\partial k_\beta}$ and $\alpha$, and $\beta$ are tensor indices. In a simpler way, the Seebeck coefficient can be expressed as

$$S = \frac{8\pi^2 k_B^2 T}{3eh^2} m^* \left(\frac{\pi}{3n}\right)^{\frac{2}{3}} \qquad (10)$$

where $m^*$ is the effective mass, and $n$ is the carrier concentration.

The value of relaxation time ($\tau$) must be computed in order to get the absolute value of these coefficients because BoltzTraP integrates electrical and electronic thermal conductivity in terms of $\tau$. We determine $\tau$ by applying the deformation potential (DP) theory to the effective mass ($m^*$) and mobility (mob$_{2D}$) of the charge carriers. The carrier mobility has been calculated using the theory.[46] Furthermore, the effective mass of the electron ($m_e^*$) and hole ($m_h^*$) has been estimated by using the parabolic curvature of the conduction (for electrons) and valence (for holes) band edges close to the Fermi level, respectively. The mathematical expression for $m^*$ is

$$m^* = \frac{\hbar^2}{\frac{\partial^2 E}{\partial k^2}} \qquad (11)$$

Moreover, the carrier mobility and relaxation time can be calculated using the following relations:

$$\text{mob}_{2D} = \frac{2e\hbar^3 C}{3k_B T |m^*|^2 E_1^2} \qquad (12a)$$

and

$$\tau = \frac{m^*}{e} \text{mob}_{2D} \qquad (12b)$$

where $k_B$ and $T$ are the Boltzmann constant and temperature, respectively, and $m^*$ is the effective mass of the charge carrier. $C$ is the elastic modulus ($C = \frac{1}{A_0} \frac{\partial^2 E}{\partial \chi^2}$, where $E$, $A_0$ and $\chi$ represent the total energy in different deformation states, lattice area at the equilibrium and the applied biaxial strain, respectively), which is determined by quadratically fitting the energy-strain data, and $E_1$ is the DP constant that reflects the strain-induced shift of the band edges (valence band maximum (VBM) for holes and conduction band minimum (CBM) for electrons).

# 3. Results and discussion

## 3.1. Structural properties

We have investigated the structural and electronic properties of monolayers BPN and I-BPN using the first principles methods. BPN and I-BPN are non-benzenoid carbons, and B–N networks are composed of octagonal, tetragonal, and hexagonal rings. Fig. 1(a) and (b) depict the atomic structure of BPN and I-BPN. Both are identical in unit cells and atomic arrangements with the only difference in their atomic entities. BPN constituents have only C atoms, whereas I-BPN has two species, B and N. Both have rectangular geometry (space group $Pmm2$; group no. 25) with six carbon atoms in BPN and three B and N atoms in I-BPN. The optimized BPN lattice parameters are $a = 3.75$ Å and $b = 4.52$ Å, which is consistent with earlier theoretical studies.[27] In contrast, the lattice parameters of the I-BPN are $a = 3.94$ Å and $b = 4.57$ Å.

The $b/a$ ratio for BPN (I-BPN) has been found to be 1.20(1.16), with a reasonably slightly larger bond length of B–B (1.62 Å) and N–N (1.55 Å) in a square motif as compared to the C–C (1.45 Å) bond length along the $x$-direction. Both monolayers have anisotropic structures due to different atomic environments, even though BPN has an identical atomic entity. This led to the expected anisotropic physical properties. There are three distinct C–C (1.45 Å, 1.40 Å and 1.44 Å) bonds in BPN, whereas I-BPN has five different bonds, three distinct B–N (1.45 Å, 1.41 Å and 1.47 Å) and two each for B–B (1.62 Å) and N–N (1.55 Å) marked as numbers in Fig. 1(a) and (b). In BPN, all of the C atoms are triangulated as expected by sp² hybridization, but the angles have different values of 90°, 110°, 125°, and 145°. These angles are distorted by a few degrees, as 90°(91.38°, 88.6°), 110°(108.48°, 110.8°), 125°(125.37°, 125.98°), and 145°(145.8°, 144.24°) for BPN (I-BPN), respectively.

## 3.2. Structural stability

The structural stability of the optimized I-BPN monolayer has been investigated through (1) cohesive and formation energy analysis, (2) phonon dispersion calculation, (3) finite temperature molecular dynamics simulation, and (4) chemisorption with O₂ molecules.

### 3.2.1. Cohesive and formation energy.

In thermodynamics, cohesive energy ($E_{coh}$) and formation energy ($E_{form}$) will be calculated to check the exothermic or endothermic synthesis process. The negative values of energies indicate the exothermic process. The $E_{form}$ is the relative energy of an I-BPN monolayer with its constituent atoms in their stable phase.

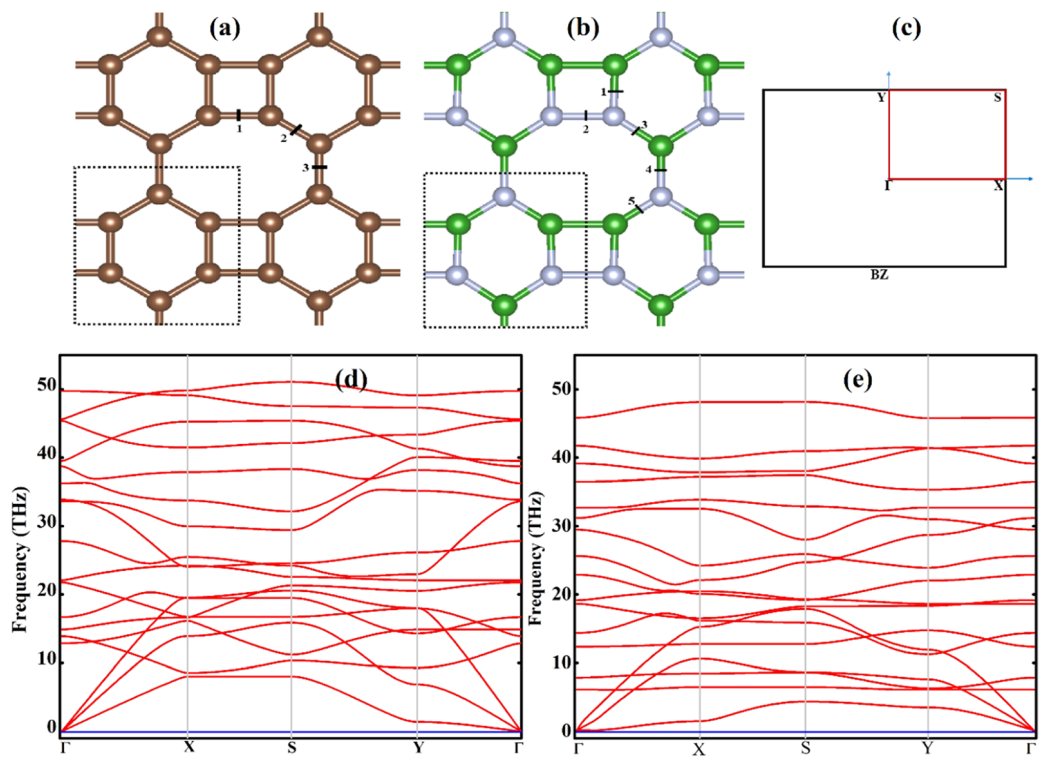

**Fig. 1** (a) and (b) Display the optimized atomic structure of BPN and I-BPN (boron in green colour and nitrogen in grey colour). The region bounded by dotted lines is the unit cell for each lattice. (c) Shows the high symmetry $K$-path in the 1st Brillouin zone (d) and (e) phonon dispersion spectra of BPN and I-BPN monolayers, respectively.

Mathematically, these energies are defined as[47-49]

$$E_{coh} = \frac{E_{monolayer} - (3E_B + 3E_N)}{6} \quad (13)$$

$$E_{form} = \frac{E_{monolayer} - (3\mu_B + 3\mu_N)}{6} \quad (14)$$

where $E_{monolayer}$ is the total energy of a unit cell of the I-BPN monolayer, and $E_B$ and $E_N$ are the energies of isolated B and N atoms. $\mu_B$ and $\mu_N$ are the chemical potentials of B and N atoms, which have been calculated from B and N-rich environments. $\mu_N = -270.22$ eV is the chemical potential of a N atom in the solid $\alpha$-N$_2$ phase and $\mu_B = -77.23$ eV is the chemical potential of a B atom in the boron-rich environment of the $\alpha$-B metallic phase. Similarly, $\mu_C = -154.86$ eV is the chemical potential of a carbon in its graphite allotrope.

According to our calculation, the $E_{coh}$ and $E_{form}$ of the I-BPN monolayer are $-7.78$ and $-0.35$ eV per atom. These negative energies show the stability of the I-BPN monolayer and it can be synthesized experimentally. Table 1 compares the $E_{coh}$ and $E_{form}$ of various 2D materials. It has been found that the magnitudes of cohesive energy and formation energy of I-BPN are remarkably equivalent to those of other experimentally synthesized monolayers.[50] The positive $E_{form}$ of graphene and BPN indicates that graphite is still more conventional for synthesis than both.

### 3.2.2. Phonon dispersion calculation.
Fig. 1(d) and (e) display the phonon spectra of BPN and I-BPN. A supercell size of $4 \times 3 \times 1$ and a $k$-point grid of $9 \times 11 \times 1$ have been used to capture the long wavelength vibration modes. The phonon spectra analysis shows no negative frequencies, indicating that the single layer of I-BPN is stable. The dispersion of phononic bands of BPN is identical to the earlier report.[51] The phonon band spectra reveal that the acoustic mode of I-BPN is less dispersive over the frequency range than the BPN, which shows that it has low phonon group velocity. Additionally, it has been found that the lowest phonon optical branches of I-BPN are slightly hybridized with the acoustic branches as compared to BPN, which means that I-BPN has more three-phonon processes and a low relaxation rate, both of which help to produce small $k_{ph}$. Between points $X$ and $S$, the transverse acoustic mode and the longitudinal acoustic mode jointly degenerate, which can greatly increase phonon scattering while reducing phonon transport, lowering the $k_{ph}$. The flatter phonon dispersion curve and the high degenerate states are important characteristics of the low phonon group velocity and small $k_{ph}$.

**Table 1** The cohesive energy, formation energy, and optimized lattice constant of graphene, h-BN, BPN and I-BPN monolayers

| Monolayer | Graphene | h-BN | BPN | I-BPN |
|---|---|---|---|---|
| Lattice constant (Å) | $a = b = 2.466$ | $a = b = 2.512$ | $a = 3.75$, $b = 4.52$ | $a = 3.94$, $b = 4.57$ |
| $E_{coh}$ (eV per atom) | $-9.16$ | $-8.72$ | $-8.69$ | $-7.78$ |
| $E_{form}$ (eV per atom) | 0.08 | $-1.23$ | 0.54 | $-0.35$ |

### 3.2.3. *Ab initio* molecular dynamics simulation.

AIMD simulations have been used to study the temperature-dependent stability of the I-BPN. Fig. 2 shows the free energy *versus* time and snapshots of I-BPN geometries at 5 ps for two different temperatures (300 K and 600 K) with a time step of 1 fs. The time step in the MD calculation has been chosen according to the maximum bond vibrations of the respective structures. For numerical stability and energy conservation, the time step ($\Delta t$) is typically chosen at least 10 orders lower than the magnitude of the fastest time step (max. vibration frequency) in the system. For example, the max. frequency of B–N bond (in-plane) vibration is $\bar{\nu} \sim 1364$ cm$^{-1}$ and the corresponding time period $\left(\tau = \dfrac{1}{\nu \bar{\nu}}\right)$ is $\sim 24.43$ fs. The time step for capturing the ionic motion is chosen as $\Delta t \sim \dfrac{\tau}{10}$, which is 2.44 fs. Still, we are safe by choosing the time step of 1 fs. It is important to note that the AIMD calculations at 600 K demonstrate that the I-BPN is thermodynamically stable because no deformation or distortion has occurred. The final snapshots of the MD simulation (shown in Fig. 2) show a negligible out-of-plane buckling at 300 K and 600 K. However, there is no significant evidence of bond breakage or structural distortion. Further, the mean and standard deviation of free energies at 300 K (600 K) are −187.62 eV (−186.98 eV), which is consistent with the literature.[52,53] The total unit cell drift (in terms of force due to MD) on the lattice constants $a$, $b$ and $c$ are 0.87, 0.52, −0.23 meV Å$^{-1}$ and 0.68, 0.63, −0.53 meV Å$^{-1}$ for 300 K and 600 K, respectively. It is worth concluding that the I-BPN monolayer is highly stable at ambient temperature.

### 3.2.4. Chemisorption of O₂ molecules.

In order to check the chemical stability of the I-BPN monolayer, the adsorption of oxygen molecules on the surface of the I-BPN monolayer has been studied with an exploration of the possibility of chemisorption or dissociation of O₂ molecules. As oxygen exists in the molecular form (O₂) in the environment, it is essential to examine the adsorption of oxygen in its molecular form on the surface of I-BPN. A supercell dimension of $2 \times 2 \times 1$ has been used to reduce the molecule–molecule interaction. The DFT-D2 method has been used to consider the vdW interaction. The adsorption of oxygen molecules has been done at six possible sites, which include the top of the centre of B–N, B–B and N–N bonds (BN-, BB- and NN-sites) and the top of the centre of square, octagon and hexagon motifs (sq-, oct-, and hexa-site). Furthermore, the adsorption energy ($E_{ads}$) of various sites in the optimized oxygen-adsorbed I-BPN monolayer has been calculated using the following relation.[54]

$$E_{ads} = E_{total} - E_{I\text{-}BPN} - E_{O_2} \qquad (15)$$

where $E_{total}$, $E_{I\text{-}BPN}$ and $E_{O_2}$ are the energies of the oxygen-adsorbed I-BPN monolayer, pristine I-BPN and isolated oxygen molecule $E(O_2)$, respectively. The optimized structures reveal that a few sites get physiosorbed, and others get chemisorbed. For illustration, the optimized geometry of BN- and hexa-sites (in Fig. S2, ESI†) is physisorbed with the O₂ at 3.2 Å and 2.8 Å distance, respectively. No chemical bond is formed between O₂ and the pristine I-BPN monolayer at these sites. On the other hand, the BB-site exhibits chemisorption, where O₂ gets chemically bonded with each boron of the B–B bond and creates two new B–O bonds with a bond length of 1.42 Å. The oxygen molecule chemisorbs at the BB-site due to its relative electron deficiency. O₂ adsorption sites closer to the B–B bond, such as NN-, sq-, and oct-sites, preferentially locked at the most favourable BB-site during optimized adsorption. Additionally, weak molecule-sheet interactions for BN and hexa-sites also suggest the possibility of surface diffusion to the BB-site, especially at room temperature. Subsequently, they may desorb after some time, or more likely, they will lock into the BB-site, potentially preventing or significantly slowing down desorption. The calculated adsorption energies of an O₂ vary between −0.62 and −0.06 eV at different sites on the I-BPN monolayer are reported in Table 2. BN- and hexa-sites have low $E_{ads}$, which is consistent with the physiorption process. Adsorption on other sites like sq-,

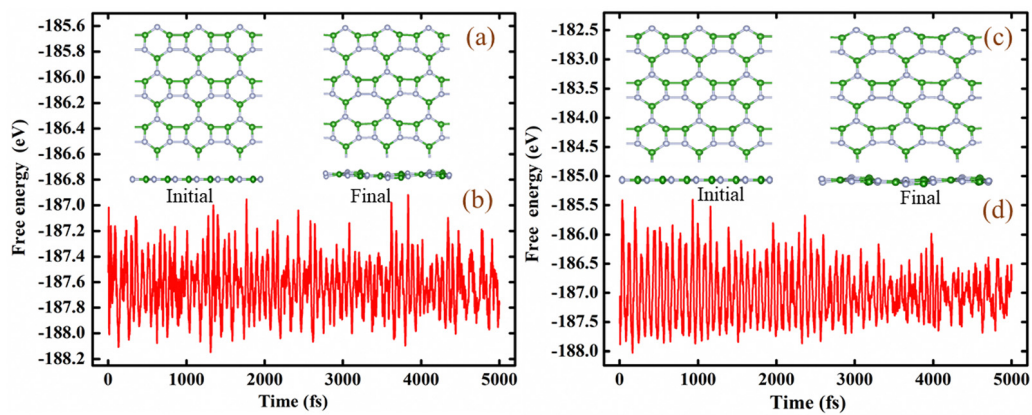

Fig. 2  (a) and (c) Show snapshots of the initial and final configuration of I-BPN and (b) and (d) display the free energy *vs.* time step for 300 K and 600 K, respectively.

**Table 2** Reports the $E_{ads}$ of an oxygen molecule on the I-BPN monolayer surface at the most favourable sites, along with the corresponding distances between the oxygen molecule and the I-BPN

| Adsorption site | BN- | BB- | Hexa- |
|---|---|---|---|
| $O_2$ $E_{ads}$ (eV) | −0.10 | −0.62 | −0.06 |
| Ads. height (Å) | 3.2 | 1.42 | 2.8 |

oct-, and NN-sites ended at the BB-site, where a substantial $E_{ads}$ value confirmed chemisorption.

Physisorption sites can potentially undergo diffusion to the chemisorption BB-site, where oxygen becomes chemically bonded with the borons of the BB-site. A distinctive out-of-plane B–O–O–B four-member motif is formed within the BB-site as oxygen bonds with both borons. The O–O bond length of the absorbed molecule increases from 1.26 to 1.38 Å, indicating a change in bond order. Additionally, the B–O bond length is measured at 1.42 Å. Further, Bader charge calculation has been employed to investigate the extent of charge transfer between the adsorbate and the I-BPN monolayer. Mathematically, it gives the charge density difference ($\rho_{net}$) of the oxygen-adsorbed system ($\rho_{total}$) and pristine I-BPN ($\rho_{I-BPN}$) and oxygen molecule ($\rho_{O_2}$).

$$\rho_{net} = \rho_{total} - \rho_{I-BPN} - \rho_{O_2} \qquad (16)$$

In Fig. 3, the charge density difference plot illustrates electron accumulation and deficiency during BB-site adsorption. The yellow region represents the electron accumulation, and the cyan colour shows the electron-deficient region in the newly bonded system. Notably, the deficiency of charge between the O atoms (cyan colour region in Fig. 3) suggests that the bond order of the O–O bond decreases. Consequently, the yellow region over the oxygen shows that it has some excess charge compared to the isolated oxygen molecule. The Bader charge analysis indicates that each oxygen atom has gained a charge of $0.65e^-$ from the I-BPN monolayer, as detailed in Table S1 of the ESI.† As a result, the I-BPN exhibits electron depletion, with both boron and nitrogen constituents acquiring positive charges. The charge deficiency is not confined solely to boron atoms but extends to the nitrogen in the adjacent B–B bonds of the monolayer. The primary contributors to the charge transfer are the B atoms (of B–B bonds) and N atoms (of N–N bonds), with positive values of $0.12e^-$ and $0.33e^-$, respectively. Other atoms also show minor charge deficiencies. This observed charge transfer results from the electronegativity difference between oxygen and the constituent atoms of I-BPN. These reported values are also consistent with the charge density difference plot depicted in Fig. 3. Hence, the chemisorption of an oxygen molecule at the BB-site gains some charge from the I-BPN sheet. Sinthika *et al.* also reported a similar charge transfer feature, where fullerene-like BN cages exhibited charge transfer to CO, $CO_2$, and $O_2$ molecules.[55]

An earlier study reported that the oxidation of graphene and h-BN is harder to do at room temperature. The defective graphene gets oxidized due to dangling bonds, or a higher temperature is required to oxidize the graphene.[56] Similarly, oxidation of h-BN is reported by annealing it in air around 850 °C, which changes its surface morphology.[57] However, the I-BPN monolayer has a B–B bond, favouring the oxygen molecule to be adsorbed through chemisorption. Chemisorption of an oxygen molecule at the BB-site of the I-BPN monolayer in an oxygen-rich environment shows chemical instability. Hence, the first principles calculation demonstrates that the chemisorption of $O_2$ on the surface of the I-BPN monolayer is thermodynamically feasible. This adsorption will also change the properties of the pristine I-BPN. The rigorous study of the dissociation of $O_2$ molecules on the surface of I-BPN (through transition state analysis) is unnecessary because feasible chemisorption already testifies that I-BPN is not chemically inert when exposed to an oxygen environment.

I-BPN exhibits an organized and self-limiting termination when exposed to $O_2$, making it a promising candidate for controlled thin-film growth *via* atomic layer deposition (ALD). ALD enables precise layer-by-layer deposition, ensuring uniformity and offering advantages like tunable thickness and enhanced environmental stability for passivated structures. This has notable implications for applications in electronic devices and other technologies exposed to external elements. In summary, the distinctive interaction between oxygen molecules and I-BPN monolayer opens avenues for controlled thin-film growth, enhancing its potential in advanced material synthesis and device fabrication.

### 3.3. Mechanical properties

For 2D systems, particularly rectangular symmetry, BPN and I-BPN have anisotropy ($a \neq b$), which implies $C_{11} \neq C_{22}$ and in order to show the mechanical stability of monolayers, their elastic constants should follow the below criteria:[58]

$$C_{11} > 0, \; C_{22} > 0, \; C_{44} > 0, \; C_{11}C_{22} > C_{12}^2$$

Additionally, Young's modulus $Y_{10}(Y_{01})$ and Poisson coefficient $\rho_{10}(\rho_{01})$ for BPN and I-BPN along the $x(y)$ direction have been

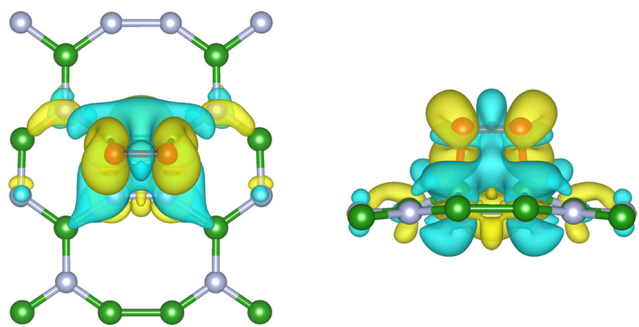

**Fig. 3** The top and side views of the charge density difference plot of the oxygen molecule adsorbed I-BPN monolayer at the BB-site in the left and right panels, respectively. The yellow region is electron accumulation, and the cyan colour is the electron depletion region.

**Table 3** compares elastic coefficients, $Y$ and $\rho$ values between the BPN and I-BPN

| Monolayer | $C_{11}$ (N m$^{-1}$) | $C_{22}$ (N m$^{-1}$) | $C_{12}$ (N m$^{-1}$) | $C_{44}$ (N m$^{-1}$) | $Y_{10}$ (N m$^{-1}$) | $Y_{01}$ (N m$^{-1}$) | $\rho_{10}$ | $\rho_{01}$ | Ref. |
|---|---|---|---|---|---|---|---|---|---|
| BPN | 292.99 | 242.63 | 94.36 | 83.40 | 256.3 | 212.24 | 0.39 | 0.32 | 59 and 60 |
| I-BPN | 219.30 | 191.82 | 70.34 | 61.19 | 193.50 | 169.25 | 0.36 | 0.32 | Present work |

calculated as;

$$Y_{10} = \frac{C_{11}C_{22} - C_{12}^2}{C_{22}}, \ Y_{01} = \frac{C_{11}C_{22} - C_{12}^2}{C_{11}}, \ \rho_{10} = \frac{C_{12}}{C_{22}} \text{ and}$$

$$\rho_{01} = \frac{C_{12}}{C_{11}}.$$

It has been found that the $Y_{10}(Y_{01})$ and $\rho_{10}(\rho_{01})$ for BPN are 256.3(212.2) N m$^{-1}$ and 0.39(0.32), respectively, which is in good agreement with earlier theoretical reports.[59,60] I-BPN also has lower $Y_{10}(Y_{01})$ and $\rho_{10}(\rho_{01})$ values than BPN, with 219.30(191.82) N m$^{-1}$ and 0.36(0.32), respectively, as reported in Table 3. The Young's modulus of I-BPN is much higher than that of black phosphorene (83 N m$^{-1}$)[61] and MoS$_2$ (123 N m$^{-1}$)[62] but lower than that of graphene ($Y_{\text{armchair}} = Y_{\text{zigzag}} = 340$ N m$^{-1}$),[63] and hexagonal BN ($Y_{\text{armchair}} = Y_{\text{zigzag}} = 275$ N m$^{-1}$).[64] These results show that I-BPN has strong mechanical properties.

Furthermore, Young's modulus $Y(\theta)$ and Poisson's ratio $\rho(\theta)$ along any arbitrary in-plane direction (where $\theta$ is the angle with respect to the $x$-direction) are determined using the formula

$$Y(\theta) = \frac{C_{11}C_{22} - C_{12}^2}{C_{11}s^4 + C_{22}c^4 + \left(\frac{C_{11}C_{22} - C_{12}^2}{C_{44}} - 2C_{12}\right)c^2 s^2} \quad (17)$$

$$\rho(\theta) = -\frac{\left(C_{11} + C_{22} - \frac{C_{11}C_{22} - C_{12}^2}{C_{44}}\right)c^2 s^2 - C_{12}(s^4 + c^4)}{C_{11}s^4 + C_{22}c^4 + \left(\frac{C_{11}C_{22} - C_{12}^2}{C_{44}} - 2C_{12}\right)c^2 s^2} \quad (18)$$

where $c = \cos\theta$ and $s = \sin\theta$ based on the calculated elastic constants. To further investigate the anisotropic mechanical properties of the BPN and I-BPN monolayers, a comparison of the in-plane Young's modulus $Y(\theta)$ and Poisson's ratio $\rho(\theta)$ has been done (using eqn (17) and (18)) between the I-BPN and BPN monolayers. In Fig. 4 (left panel), the 2D polar plot of Young's modulus (as a function of $\theta$) first goes from a maximum of 193.50 N m$^{-1}$ in the $x$-direction ($\theta = 0°$) to a minimum of 169.25 N m$^{-1}$ in the $y$-axis ($\theta = 90°$), and then gradually rises and achieves a maximum of 193.50 N m$^{-1}$ at $\theta = 180°$. The 2D polar plot reveals an oval curve, indicating that Young's modulus, like BPN, is anisotropic across the in-plane of the I-BPN monolayer. Further, a positive Poisson's ratio in Fig. 4 (right panel) for BPN and I-BPN implies a tendency to expand or contract in the opposite direction of a compressive or tensile strain.

Our calculations reveal that the BPN and I-BPN structures are almost identical in terms of their structural stability, which can be due to their equal bond strength. The average bond energy of the C–C (sp$^2$ hybridized) and B–N bonds is 6.98 eV per atom and 6.45 eV per atom, respectively.[65,66] The similar stability of BPN and I-BPN structures can be attributed to bond strength, as evidenced by their nearly identical formation energies, mechanical strength, and temperature sustainability.

### 3.4. Electronic properties

The electronic band structures of BPN and I-BPN along the high symmetry path are shown in Fig. 5. It is evident that BPN shows metallic behaviour because of crossing a few bands at the Fermi level, which is consistent with the previous study.[27] However, I-BPN is an indirect bandgap ($E_g = 1.88$ eV) semiconductor with the valence band maximum (VBM) at $Y$ and conduction band minimum (CBM) at $S$. Along the high symmetry $Y-\Gamma$ path, BPN shows the linear crossing of bands next to the Fermi level that is slightly tilted to the side. The solid-state system having this tilted Dirac cone is defined as a system where the effective space-time is non-Minkowski.[67] The tilted Dirac cone has been seen in various Dirac/Weyl materials.[68–70] BPN and I-BPN are entirely different in terms of their electronic properties. The electronic properties sharply depend on the valence electrons of constituent atoms in crystal structures and the nature of the bonding between the atoms. Furthermore, whether a material is conducting (metal) or non-conducting (semiconductor or insulator) depends on the presence or absence of free or delocalized electrons. BPN is metallic because it solely consists of carbon (C) atoms covalently bonded with three other C atoms and the $p_z$ electrons of each C atom are delocalized across the 2D network. In I-BPN, the B–N bond is partially ionic due to the difference in electronegativity of boron and nitrogen. The electronegativity difference between B and N is 1.0 Pauling scale, making the B atom an electron donor and N an electron acceptor. The partially ionic B–N bond in I-BPN reduces its covalent nature due to charge localization on atomic sites. The local charge distribution in BPN and I-BPN has been calculated in terms of electron localization function (ELF) with an iso-surface of 0.02 a.u. The ELF plots are shown in Fig. S3 (ESI†). The ELF plots clearly show that charges are distributed uniformly over the C–C bond in the BPN structure, whereas the charge envelope is more polarised towards the nitrogen site in I-BPN. In general, charge localization leads to the insulating nature of materials.[71,72] That is why I-BPN is a semiconductor and BPN is a metal. Further, the projected density of states of the I-BPN monolayer (in Fig. S4, ESI†) shows that the $p_z$ orbitals of N atoms contribute the most to the valence band maximum (VBM) in I-BPN, whereas the $p_z$ orbitals of B atoms contribute the most to the conduction band minimum (CBM). Additionally, the VBM and CBM charge density plots support the relative contributions of B and N, as shown in Fig. S5 (ESI†). In the meantime, the thermoelectric parameters like the Seebeck coefficient strongly depend on the dispersion of electronic bands around the Fermi energy of semiconductor materials. Hao *et al.* have demonstrated that the non-dispersive valence/conduction band near the Fermi energy is responsible for a high Seebeck coefficient.[73] The valence band of I-BPN is nearly flat along the $Y$–$\Gamma$ path, indicating a high band effective mass of the carrier that helps to enhance the Seebeck coefficient.

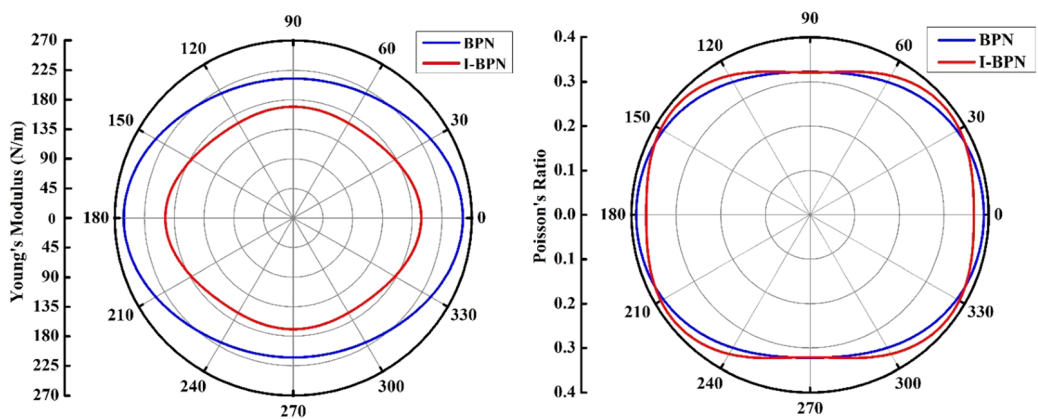

**Fig. 4** Polar diagram for Young's modulus $Y$ (left) and Poisson's ratio (right) of BPN (blue) and I-BPN (red) monolayers, respectively.

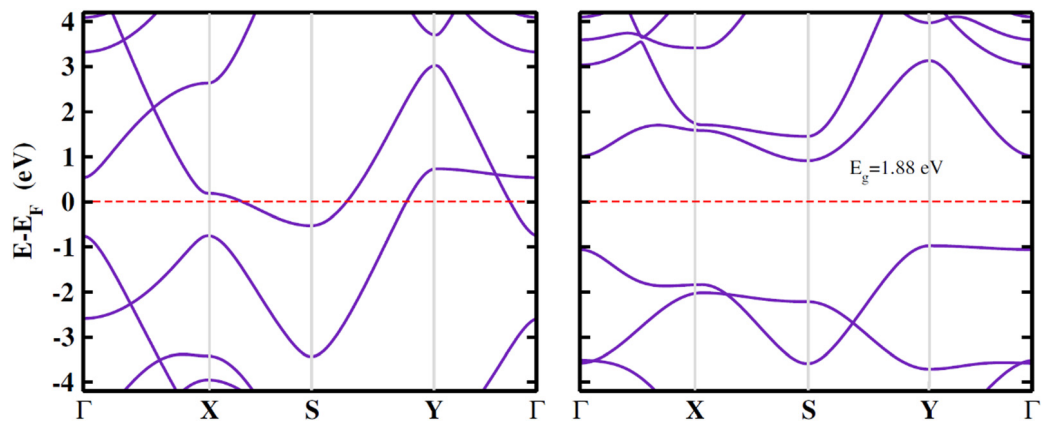

**Fig. 5** Electronic band spectra of BPN (left) and I-BPN (right) monolayers with the Fermi energy (red dashed line) scaled to zero.

Because of this unique flat band, I-BPN could be an efficient thermoelectric material.

### 3.5. Lattice thermal conductivity

We use the Phonopy and ShengBTE packages to investigate the relationship between the phonon thermal conductivity ($k_{ph}$) of the single-layer of BPN (I-BPN) and temperature. Normally, the thermal conductivity of the lattice follows a $T^{-1}$ trend with temperature. The temperature-dependent lattice constants are not included in the BTE solution in this study, assuming that the thermal expansion of these lattices at high temperatures has no significant effect on the phononic properties. For BPN and I-BPN monolayers, we observed nearly identical temperature power factors in both the $x$- and $y$-direction.

Fig. 6 depicts the $k_{ph}$ of BPN and I-BPN, which shows a typical temperature-dependent behaviour for the semiconductor family[74] in the 200–1000 K range. The $k_{ph}$ of BPN is anisotropic along the $x(y)$-axis due to its different atomic environments. At 300 K, it is revealed that the $k_{ph}$ of BPN is 398 W m$^{-1}$ K$^{-1}$ and 187 W m$^{-1}$ K$^{-1}$ along the $x$- and $y$-axes, which is consistent with previous studies.[39,75] Furthermore, at room temperature, the anisotropic $k_{ph}$ of I-BPN along the $x$- and $y$-axes are 21.4 W m$^{-1}$ K$^{-1}$ and 19.8 W m$^{-1}$ K$^{-1}$, respectively. The $k_{ph}$ of BPN has been found to be ten times larger than that of the I-BPN, which would be expected because hexagonal graphene and h-BN monolayers have been reported with an identical anisotropic difference in $k_{ph}$.[76–80]

### 3.6. Thermoelectric transport properties

The ideal thermoelectric material should resemble a perfect single crystal in terms of electrical characteristics and glass in terms of thermal features. As metals (insulators) have good (bad) electrical and thermal conductivity, these may not be suitable for thermoelectric study. On the other hand, a semiconductor would be a good choice for thermoelectric investigation since its characteristics lie between metals and insulators. BPN is not good for thermoelectric studies because of its metallic behaviour as seen in the electronic band spectrum. In contrast, I-BPN is a semiconductor with an indirect bandgap of 1.88 eV; it is worth studying its thermoelectric properties. Meanwhile, the calculated lattice thermal conductivity of an I-BPN monolayer is also in the significant range, indicating that I-BPN may be a good thermoelectric material.

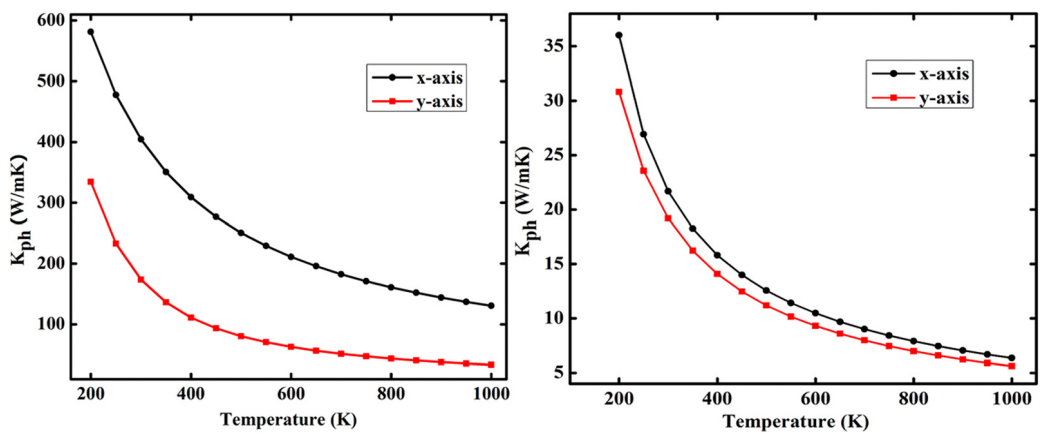

**Fig. 6** Anisotropic phonon thermal conductivity *vs.* temperature plots of BPN (left) and I-BPN (right) monolayers.

**Table 4** The charge carrier and their corresponding effective mass, deformation potential, mobility, and relaxation time along the *x*- and *y*-axes have been calculated. The mobility and relaxation time are calculated at 300 K

|  | Charge carrier | $m_e^*(m_e)$ | $E_1$ (eV) | $C^{2D}$ (N m$^{-1}$) | mob. (cm$^2$ V$^{-1}$ s$^{-1}$) | $\tau$ (fs) |
|---|---|---|---|---|---|---|
| I-BPN(x) | Electron | 0.74 | −1.94 | 108.22 | 746.15 | 313.51 |
| I-BPN(y) | Electron | 0.51 | 4.41 | 94.68 | 264.10 | 76.62 |
| I-BPN(x) | Hole | 0.70 | 3.06 | 108.22 | 340.54 | 135.39 |
| I-BPN(y) | Hole | 0.67 | 4.04 | 94.68 | 186.89 | 71.12 |

Under the rigid bands and constant relaxation time approximation, the BoltzTraP program integrates the electronic Boltzmann transport equations. It is assumed that the electronic band structure of the host material remains the same with doping; only the chemical potential changes with the doping and temperature. In energy-independent relaxation time, the transport coefficients, such as the electrical and electronic thermal conductivity, are derived in terms of relaxation time, which explicitly depends on temperature and energy. These approximations are intensively verified theoretically and experimentally for renowned thermoelectric materials like PbTe[81] and HfCoSb.[82]

The semi-classical transport Boltzmann equations give thermoelectric coefficients such as the Seebeck coefficient ($S$), which is independent of the relaxation time ($\tau$), whereas electrical conductivity ($\sigma$) and electronic thermal conductivity ($\kappa_e$) are linearly dependent on $\tau$. Therefore, a suitable value of $\tau$ should be chosen to find the absolute value of $\sigma$ and $\kappa_e$. As a matter of fact, $\tau$ of a material depends on the carrier concentration and temperature. So far, the experimental measurement is the most effective way to choose the $\tau$. To the best of our knowledge, no experiments have been carried out to measure the relaxation time of charge carriers in I-BPN. The theoretical calculations are the only way to find the value of $\tau$ for our systems. Therefore, as discussed earlier, we have chosen the deformation potential (DP) theory to calculate the relaxation time. From the above electronic band structure, the effective mass is determined for the n-type and p-type carriers of I-BPN. The calculated elastic modulus, carrier mobility (at $T$ = 300 K), relaxation time (at $T$ = 300 K), and the DP constant values for the electrons and holes of I-BPN are shown in Table 4. The DP theory is good for determining the order of $\tau$ but does not provide an exact value because it only considers longitudinal acoustic phonon scattering. This theory does not consider the scattering of charge carriers with transverse acoustic phonons, optical phonons and other carriers.[83] As $\tau$ is anisotropic and carrier dependent, the absolute values of electronic transport coefficients have been calculated using the mean value of $\tau$ (149.21–150 fs) calculated at room temperature. In similar 2D materials, the values of $\tau$ have been reported in the same order of $10^{-13}$ s.[84–87]

Further, the $\tau$ value at various temperatures has been estimated by the relation $\tau_T = \dfrac{300 \times \tau_{300}}{T}$, where $\tau_T$ is the relaxation time at temperature $T$.[88,89] This relation could also be verified by using eqn (11).

To obtain more insight, Fig. 7(a)–(d) display the Seebeck coefficient ($S$), electrical conductivity ($\sigma$), electronic thermal conductivity ($\kappa_e$) and electronic figure of merit ($ZT_e$) as a function of the chemical potential ($\mu$) at three different temperatures, 300 K, 600 K and 900 K, for both transport (*x*- and *y*-axis) directions.

It has been shown that $\mu$ is positive for n-type doping and negative for p-type doping. $S$ is symmetric about $\mu$ = 0 for the I-BPN monolayer for dilute doping. As shown in Fig. 7(a), the $S$ values are negligible anisotropic. At 300 K, the I-BPN monolayer has the maximum Seebeck coefficient, 0.00289 V K$^{-1}$, which decreases with increasing temperature. The bipolar effect occurs as the temperature increases, causing an increase in carrier concentration and a corresponding decrease in the Seebeck coefficient.[90] Further, the $S_{max}$ of the I-BPN monolayer is slightly higher than that of the semiconducting allotropes of carbon like graphdiyne (0.000248 V K$^{-1}$) and γ-graphyne (0.000260 V K$^{-1}$)[91] and other thermoelectric monolayers like SnS (0.00145 V K$^{-1}$), arsenene (0.00118 V K$^{-1}$), and phosphorene (0.0013 V K$^{-1}$).[86,92] This high Seebeck coefficient value of I-BPN is due to the non-dispersive nature of valence bands along the $Y$–$\Gamma$ path. The flat or non-dispersive band shows a high effective mass of the charge carrier, which is directly proportional to the Seebeck coefficient, as seen in eqn (10). In addition, the Seebeck coefficient ($S$) is higher for low carrier concentration and it begins to decrease with increasing carrier concentration. The carrier concentration increases with temperature and doping effects. The calculated electrical and electronic thermal conductivities, both scaled by $\tau$ (that has been estimated by DP theory), are shown in Fig. 7(c) and (b), respectively. $\sigma$ exhibits strong anisotropic behaviour along the $x$- and $y$-axes, yielding a typical response with $\mu$ and temperature. Generally, the materials with narrower gaps have higher conductivities, and I-BPN shows a low $\sigma$ because of the wider band gap. Similarly, electronic thermal conductivity ($\kappa_e$) is highly influenced by temperature and chemical potential. The $\kappa_e$ is non-tunable since lowering the electronic component will automatically lower the $\sigma$ as well because the same charge carrier drives both conductivities. So, the low lattice thermal conductivity is important for a high $ZT$ value. The electronic contribution of thermal conductivity as a function of chemical potential at 300 K, 600 K, and 900 K is shown in Fig. 7(b). The $\kappa_e$, at 300 K, is lower than that of other thermoelectric materials. Further, it has been found that the overall variation of $\kappa_e$ with temperature and doping looks nearly similar to the variation of $\sigma$ with temperature and doping. At room temperature, $\kappa_e$ along the $x$-axis is higher than that along the $y$-axis, showing its strong anisotropic behaviour. Furthermore, the thermoelectric performance of a material is examined by a figure of merit ($ZT = S^2\sigma T/\kappa_e + \kappa_{ph}$). Firstly, we consider only the electronic part, $ZT_e = S^2\sigma T/\kappa_e$. In Fig. 7(d), $ZT_e$ exhibits two amplified peaks in dilute n- and p-type doping regions. It has been observed that the maximum of $ZT_e$ appears in that region of $\mu$, where the $S$ is high and $\kappa_e$ is low for a particular temperature value. Further, $ZT_e$ slightly decreases from 0.99(0.98) to 0.98(0.96) along the $x(y)$ direction as $T$ goes from 300 K to 900 K.

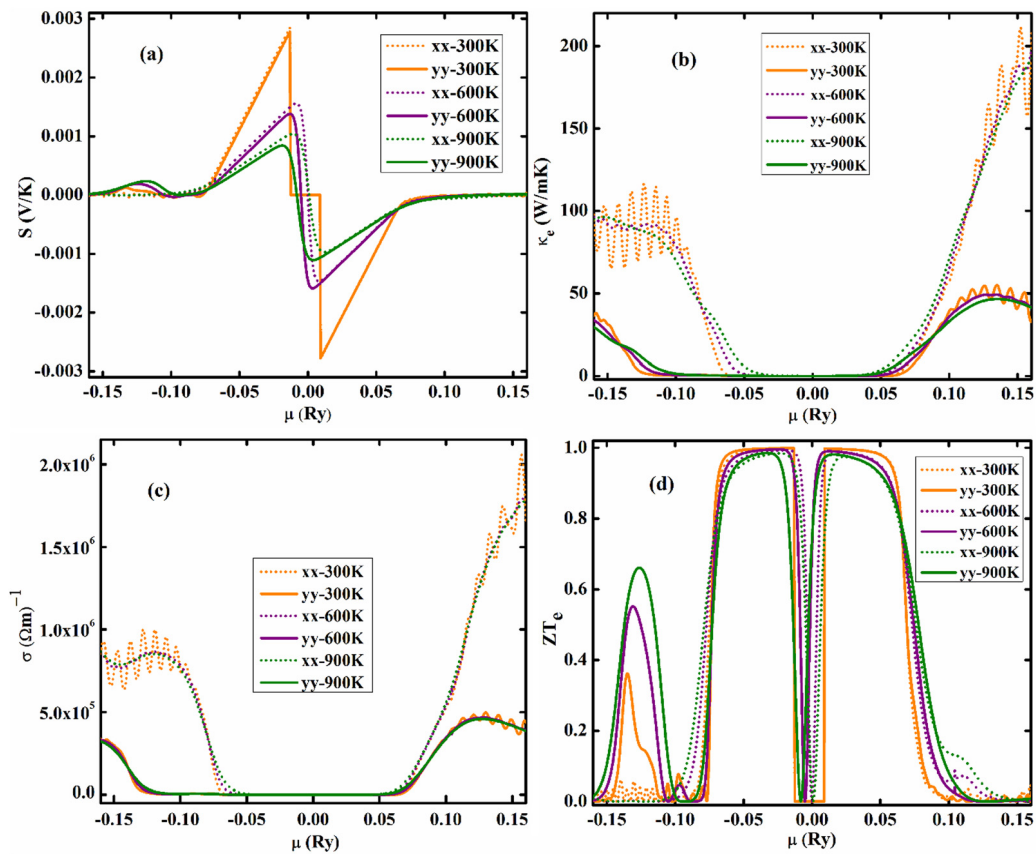

**Fig. 7** Anisotropic thermoelectric parameters: (a) Seebeck coefficient, (b) thermal conductivity, (c) electrical conductivity, and (d) electronic figure of merit as a function of chemical potential for the I-BPN monolayer.

Additionally, another thermoelectric quantity, the power factor (PF = $S^2\sigma$), must be defined. PF in thermoelectric power generation indicates how much energy is produced at a given

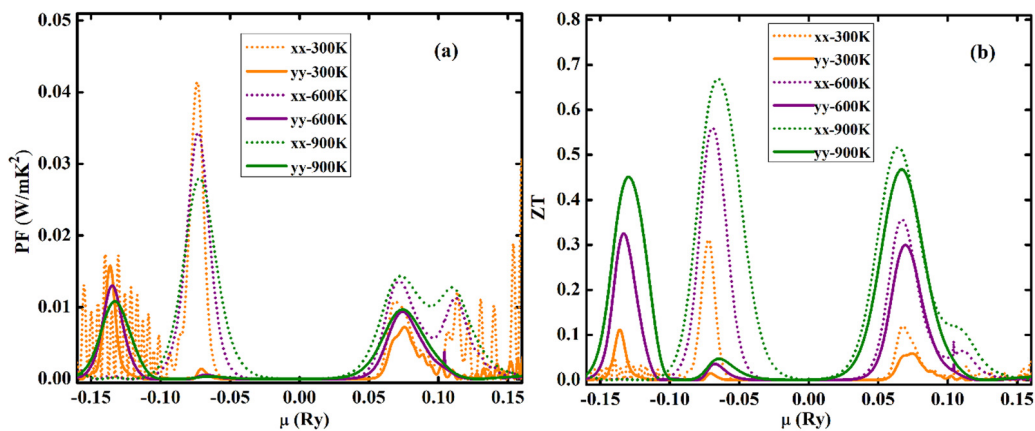

**Fig. 8** (a) Thermoelectric power factor and (b) total figure of merit as a function of chemical potential in the I-BPN monolayer.

temperature. It conveys the role of fermions as energy carriers in thermoelectric power generation. In Fig. 8(a), the peak value of PF is highest at 300 K and decreases with temperature in both directions. This decrease is reasonable because the peaks appear where $\sigma$ starts to increase and $S$ follows the downfall, especially for the p-type doping. The maximum PF value for p-type doping is 0.057 W m$^{-1}$ K$^{-2}$ at 300 K along the $x$-axis. In addition, the peak value of PF along the $y$-axis has been observed at higher chemical potential, and a relatively low value of 0.015 W m$^{-1}$ K$^{-2}$ shows strong anisotropy in PF. Since $S$ is symmetric for both directions, PF is highly anisotropic because of $\sigma$. In contrast, the peak values of PF in the n-type doping zone are low compared to p-type doping because the significant values of $S$ and $\sigma$ do not meet each other, as $\sigma$ starts increasing when $S$ trends to zero. The peak value of PF is comparable to that of graphyne,[91] arsenene[10] and SnS monolayers.[92] Fig. 8(b) depicts the total $ZT$ (considering both $\kappa_e$ and $\kappa_{ph}$), and it has been found that the total contribution in $\kappa$ is dominated by the lattice part at low temperatures; meanwhile, at higher temperatures, $\kappa_e$ suppresses the $\kappa_{ph}$. At low temperatures, $\kappa_{ph}$ is relatively high due to the weak anharmonicity in the covalent bonds and vibrations in the I-BPN monolayer. In the meantime, the $ZT$ peak value is temperature-dependent and rises to its highest at 900 K. In Fig. 8(b), it has been found that the $ZT$ peak value is 0.68(0.46) at 900 K along the $x(y)$ direction in the p-type doping region, whereas in the n-type doping region, the $ZT$ value is 0.56(0.48) in the $x(y)$ direction. These high peak values of $ZT$ along the $x$-axis are mainly caused by the dominance of electrical conductivity over other directional dependent thermoelectric coefficients ($S$, $\kappa_{ph}$ and $\kappa_e$). Further, the I-BPN monolayer is more likely to be p-type thermoelectric as the $ZT$ peak value for p-type doping is higher than that for n-type at 900 K. In summary, the combination of a high Seebeck coefficient and low thermal conductivity makes I-BPN a better thermoelectric material compared to other B–N compositions.

## 4. Conclusions

To summarise our results, the first principles computations have been used to investigate the structural, elastic, electronic, and lattice transport features of BPN and I-BPN. It has been discovered that the BPN and I-BPN are comparably stable in terms of formation energy and phonon dispersion spectra. Chemical inertness has been tested by adsorbing an oxygen molecule on the surface of the I-BPN monolayer, revealing a preference for oxygen chemisorption on B–B bonds. We have also noticed that the Young modulus of I-BPN is slightly lower than that of experimentally synthesized BPN. The obtained values of Young modulus and Poisson's ratio indicate that both are mechanically stable. The nearly equal bond strength of C–C and B–N in BPN and I-BPN monolayers renders both equally stable. The lattice thermal conductivity of I-BPN is ten times lower than that of the BPN monolayer, which is not surprising as the same order of difference between graphene and the h-BN monolayer is already reported. Furthermore, the electronic properties reveal that BPN is metallic and I-BPN is semiconducting in nature. This difference is due to the delocalization and localization of charge in BPN and I-BPN, respectively. In thermoelectric studies, the Seebeck coefficient has a significant value, 0.00289 V K$^{-1}$ at 300 K, in both the electron- and hole-doped regions. The power factor reveals that the p-type doping shows a peak value of 0.057 W m$^{-1}$ K$^{-2}$ at 300 K along the $x$-axis compared to the 0.0128 W m$^{-1}$ K$^{-2}$ in the n-type region. Moreover, $ZT_e$ shows a peak value of 0.99 and 0.98 in the p-type and n-doping regions. In addition, the total $ZT$ peak value is 0.68(0.46) along the $x(y)$ direction at 900 K. The $ZT$ peak in the p-type region is 0.68 at 900 K along the $x$-axis, indicating that electron-deficient doping is more suitable for I-BPN. Our study is a modest attempt to highlight environment-friendly thermoelectric materials.

## Author contributions

A. K.: calculation, methodology, analysis, draft writing and editing; P. S.: analysis and draft editing; P. P.: supervision (leading).

## Conflicts of interest

There are no conflicts to declare.

## Acknowledgements

AK thanks University Grants Commission (UGC), New Delhi, Government of India, for financial support in the form of a Senior Research Fellowship (DEC18-512569-ACTIVE). PS acknowledges the DST INSPIRE (IF190005) Government of India, for financial assistance. PP thanks DST-SERB Government of India, for ECRA project (ECR/2017/003305).

## Notes and references